\documentclass{Interspeech}
\usepackage[table]{xcolor}
\usepackage{subcaption}
\usepackage{svg}
\usepackage{url}
\usepackage{multirow}
\usepackage{colortbl}
\usepackage{subcaption}
\usepackage{float}
\restylefloat{table}
\usepackage{lipsum}
\usepackage[utf8]{inputenc}
\usepackage{ragged2e}
\usepackage{multicol}
\usepackage{cite}
\usepackage{amsmath,amssymb,amsfonts}
\usepackage{algorithmic}
\usepackage{graphicx}
\usepackage{pifont}
\usepackage{textcomp}
\usepackage{hyperref}

\interspeechcameraready

\title{\textit{Dhvani}: A Weakly-supervised Phonemic Error Detection and Personalized Feedback System for Hindi}

\author[affiliation={1}]{Arnav}{Rustagi}
\author[affiliation={1}]{Satvik}{Bajpai} 
\author[affiliation={1}]{Nimrat}{Kaur} 
\author[affiliation={1}]{Siddharth}{Siddharth}

\affiliation{HTI Lab}{Plaksha University}{Mohali, India}

\email{arnav.rustagi@plaksha.edu.in, satvik.bajpai@plaksha.edu.in, nimrat.kaur@plaksha.edu.in, siddharth.s@plaksha.edu.in}


\usepackage{comment}

\begin{document}

\maketitle

\begingroup
\renewcommand\thefootnote{*}
\endgroup

\begin{abstract}
    
    Computer-Assisted Pronunciation Training (CAPT) has been extensively studied for English. However, there remains a critical gap in its application to Indian languages with a base of 1.5 billion speakers. Pronunciation tools tailored to Indian languages are strikingly lacking despite the fact that millions learn them every year. With over 600 million speakers and being the fourth most-spoken language worldwide, improving Hindi pronunciation is a vital first step toward addressing this gap. This paper proposes 1) \textit{Dhvani}---a novel CAPT system for Hindi, 2) synthetic speech generation for Hindi mispronunciations, and 3) a novel methodology for providing personalized feedback to learners. While the system often interacts with learners using Devanagari graphemes, its core analysis targets phonemic distinctions, leveraging Hindi's highly phonetic orthography to analyze mispronounced speech and provide targeted feedback.
\end{abstract}

\section{Introduction}
India's linguistic diversity, while a cultural asset, presents significant challenges for spoken communication and linguistic cohesion within the country. Indigenous languages remain widely in use and cannot be replaced by a single lingua franca, often necessitating individuals to learn multiple languages throughout their lives due to migration for work or education. Mastering the pronunciation and cultural nuances of a new language can be a formidable obstacle. These speech barriers can hinder clear verbal communication in unfamiliar settings, potentially affecting social integration, access to opportunities, and overall national cohesion.\cite{census, NEP2020} 

Effective communication in a second language relies heavily on confident speech, which in turn is only achievable through accurate pronunciation. \cite{bakar2015importance} In the world's most populous country, providing pronunciation training through language experts demands vast human resources and lacks scalability. Therefore, Computer-Assisted Pronunciation Training (CAPT) systems for Indian languages may offer a promising and practical solution to support second language acquisition at scale. 

With over 600 million speakers, Hindi is the fourth most spoken language globally, with millions more aspiring to learn it every year, as it is one of India's two official languages. Additionally, unlike English's many-to-many phoneme-grapheme mapping, Hindi mostly has a one-to-one correspondence, exhibiting high grapheme-to-phoneme consistency. Yet, the development of a CAPT system for Hindi has been underexplored.

The proposed CAPT system, \textit{Dhvani} (literally meaning ``sound") with its capability to provide personalized pronunciation feedback is a first-of-its-kind for Hindi. \textit{Dhvani} uses a weakly-supervised encoder-decoder structure. The encoder employs a Recurrent Convolutional Neural Network (RCNN) for feature extraction, while the decoder uses a shared Attention-based Recurrent Neural Network (ARNN) to model phoneme sequences. \cite{dk} The scarcity of Hindi mispronounced speech data---crucial for developing a robust CAPT system---is addressed by speech synthesis. This approach not only overcomes data limitations but also ensures diversity in speech samples, making it an inclusive and representative system. \cite{korzekwa2022computer} 

Our approach leverages the predictable relationship between written symbols (graphemes) and sounds (phonemes) in Hindi. This simplifies the representation of phonemes and enables the provision of providing comprehensive personalized feedback on pronunciation errors. Our feedback mechanism provides detailed instructions on tongue position, lip movement, and teeth placement, supplemented by visual aids in the form of tongue diagrams. \cite{malviya, PATIL2016202} 

\section{Literature Survey}
\label{sec:litsurvey}

\begin{table*}[ht]
    \centering
    \renewcommand{\arraystretch}{1.4} 
    \begin{tabular}{|p{2.7cm}|p{2.7cm}|p{1.3cm}|p{4.1cm}|p{2.3cm}|p{1.2cm}|}
        \hline
        \rowcolor{gray!20} \textbf{Research Work} & 
        \textbf{Methodology} & 
        \textbf{Language} &
        \textbf{Dataset} &
        \textbf{Metrics} & 
        \textbf{Feedback} \\ \hline
        Bhatt et al.\cite{bhatt2020confusion} & 
        HMM &
        Hindi & 
        TIMIT &
        Accuracy: 72.2\% &
        \ding{55} \\ \hline
        Dash et al.\cite{dash2018automatic} & 
        SD-DNN &
        Hindi & 
        Original Speech Corpus\cite{dash2018automatic} &
        PER: 27\% &
        \ding{55} \\ \hline
        Pradeep et al. \cite{pradeep2016deep} & 
        HMM-DNN &
        Kannada & 
        Kannada Speech Corpus\cite{shridhara2013development} &
        PER: 39.9\% &
        \ding{55} \\ \hline
        Kotwal et al.\cite{kotwal2010bangla} & 
        MLN-HMM &
        Bangla & 
        Bangla Speech Corpus\cite{kotwal2010bangla} &
        Accuracy: 47.5\% &
        \ding{55} \\ \hline
        Mukherjee et al.\cite{mukherjee2017read} & 
        MLP-based classifier &
        Bangla &
        Bangla Phoneme Database\cite{mukherjee2017read}  &
        Accuracy: 98.4\% &
        \ding{55} \\ \hline
        Zhang et al.\cite{zhangetalforresults} & DNN\_DNN\_AGP & Mandarin & CCTV \& PSC-G1-112 & F1-score: 62.48\% & \ding{55} \\
         \hline
         Baranwal et al.\cite{nehab} & CTC-ATT & English & TIMIT \& L2-ARCTIC\cite{zhao2018l2} & F1-score: 56.08\% & \ding{55} \\
         \hline
         Korzekwa et al.\cite{dk} & WEAKLY-S & German \& Italian & Isle Corpus & F1-score: 51.53\% & \ding{55}\\
         \hline
         Korzekwa et al.\cite{dk} & WEAKLY-S & Polish & GUT Isle Corpus & F1-score: 52.56\% & \ding{55} \\
         \hline
        \textbf{\textit{Dhvani} (This work)} & 
        \textbf{RCNN-ARNN} &
        \textbf{Hindi} & 
        \textbf{MUCS\cite{mucs} + Speech Synthesis} &
        \textbf{F1-score: 82\%} &
        \textbf{{\ding{51}}} \\ \hline
    \end{tabular}
    \caption{Existing literature on CAPT for Indian and other Indo-European languages. PER stands for Phoneme Error Rate.}
    \label{tab:literature}
\end{table*}

CAPT has made significant strides in second language acquisition, yet existing models struggle with detecting pronunciation errors accurately. \cite{leung2019cnn, zhang2022l2} Currently, CAPT systems for Indian languages remain largely unexplored, mostly focusing on phoneme recognition tasks for detecting mispronunciation with limited success (Table~\ref{tab:literature}). Their key limitation lies in the paucity of mispronounced speech data---a prerequisite to developing CAPT systems. The literature proposes three notable speech synthesis techniques to address this issue:

\begin{enumerate}
    \item Phoneme-to-Phoneme (P2P): Perturbs phonetic transcriptions to simulate errors but lacks variability in prosody and is unable to generate new speech signals.\cite{zhang2022l2}
    \item Text-to-Speech (T2S): Uses neural TTS to convert phoneme sequences into speech, offering improved variability, but producing the same output for identical inputs.\cite{liu2024controllable, shah2021non}
    \item Speech-to-Speech (S2S): Introduces mispronunciations into aligned mel-spectrograms, maintaining prosody and voice timbre while generating diverse speech.\cite{korzekwa2022computer} 
\end{enumerate}

The WEAKLY-S model, proposed by Korzekwa et al., used the robust S2S technique and bypassed the need for detailed phonetic transcription. It could work on word-level mispronunciation detection with both annotated L1 (native speech) data and unannotated L2 (non-native speech) data. This significantly reduced the data annotation burden, addressing a key challenge in developing robust CAPT systems.
\cite{korzekwa2021mispronunciation,zhang2022l2} The WEAKLY-S model uses two interconnected sub-networks:
\begin{enumerate}
    \item Mispronunciation Detection Network (MDN): This network detects pronunciation errors at the word-level by comparing the input speech signal with canonical phonemes.
    \item Phoneme Recognition Network (PRN): This network recognizes phonemes in the speech signal, enhancing the model’s robustness by processing both L1 and L2 data.
\end{enumerate}

Both networks share an RCNN encoder that processes mel-spectrogram representations and an attention-based RNN decoder. This shared architecture serves as a regularizer, preventing overfitting while leveraging features from both phoneme recognition and mispronunciation detection tasks.\cite{zhang2022l2} This innovative design efficiently detects mispronunciations without requiring detailed phonetic annotations.\cite{dk,Korzekwa_2022} The architecture effectively addresses two key challenges in CAPT systems: the scarcity of mispronounced speech data and the difficulty in transcribing mispronounced speech.\cite{zhang2022l2}

Unlike the original WEAKLY-S model, our approach introduces several key modifications based on extensive experimentation. Most notably, we removed the shared decoder architecture after observing that it added unnecessary complexity to Hindi speech. We noticed that using separate decoders achieved a higher performance compared to shared decoders, likely due to Hindi's highly consistent phonetic structure requiring less cross-task regularization. A crucial discovery was the importance of maintaining prosodic variations---when removed, the model's performance dropped drastically, as these variations are essential for distinguishing between phonetically similar but prosodically distinct Hindi word pairs. To preserve these crucial features, we incorporated energy adjustments (±5dB) and speed modifications (±10\%) in our augmentation pipeline. Additionally, we found that replacing Location-sensitive Attention with Multi-head Attention in the decoder substantially improved the model's performance, as it better captured Hindi-specific phonetic dependencies and prosodic patterns. These architectural changes, validated through rigorous ablation studies, demonstrate the importance of sufficiently modifying the WEAKLY-S architecture to account for language-specific characteristics while preserving critical prosodic information.
\begin{figure*}[t]
    \centering
    \begin{subfigure}[b]{0.9\textwidth}
        \centering
        \includegraphics[width=\textwidth]{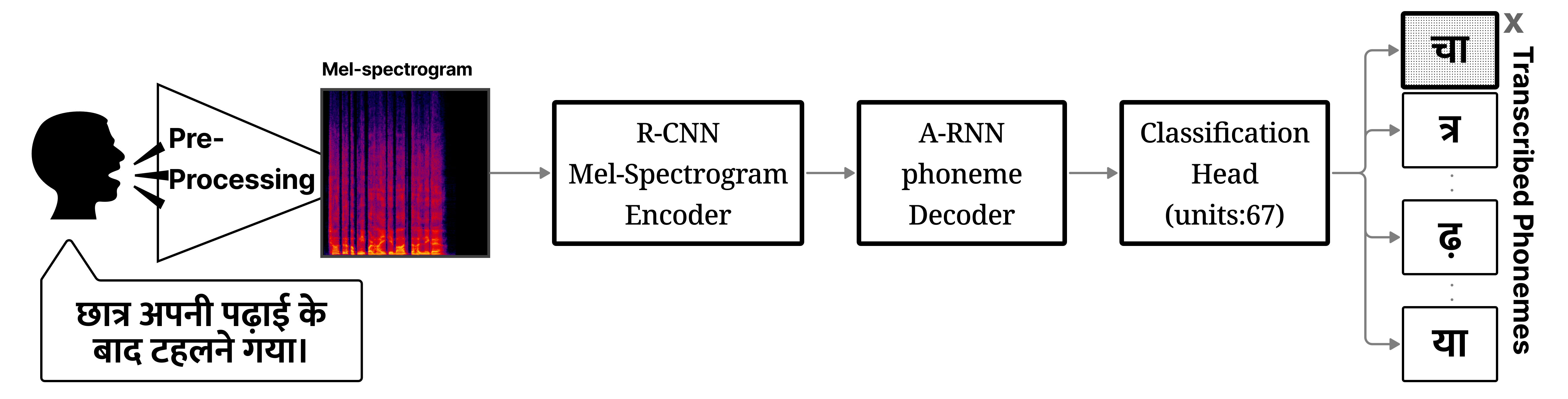}
        \caption{Model architecture: Input speech preprocessing, R-CNN Mel-Spectrogram Encoder, A-RNN phoneme Decoder, and Classification Head (67 units for Hindi phonemes). An example for Hindi language input sentence in the Devanagari script.}
    \end{subfigure}
    
\begin{subfigure}[b]{0.75\textwidth}
    \centering
    \includegraphics[width=\textwidth]{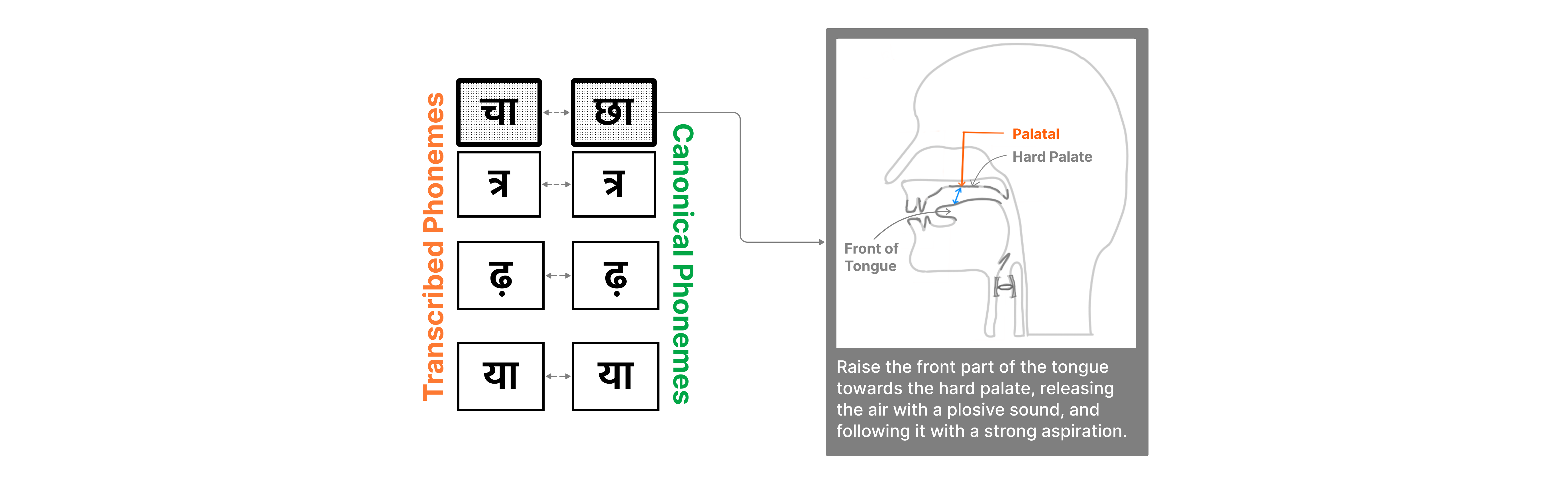}
    \caption{Feedback mechanism: Comparison of transcribed and canonical phonemes. Detailed articulatory feedback with tongue diagram for mispronounced phonemes. Example: unvoiced palatal aspirated affricate.}
\end{subfigure}

    \caption{\textit{Dhvani} system overview: (a) Model architecture and (b) Personalized feedback mechanism}
    \label{fig:dhvani-model-and-feedback}
\end{figure*}

\section{Methodology}
\label{sec:dataset}

\subsection{Dataset}
The MUCS dataset (for L1 Hindi Speech), a subset of the Multilingual and Code-switching ASR Challenge dataset curated by the MUCS organization and available on OpenSLR, was utilized for this research.\cite{mucs} The dataset consists of telephone-quality Hindi speech recordings collected via the Mobile Vaani platform from speakers across India, with corresponding transcriptions provided by crowd workers. The audio files are sampled at 8 kHz with 16-bit encoding, each lasting 6-13 seconds and containing a single sentence. Multiple speakers recorded for the same sentence, creating diverse pronunciation samples. While primarily designed for Automatic Speech Recognition (ASR), the dataset's characteristics make it a valuable resource for this research.

\subsection{Preprocessing and Speech Synthesis}
To augment our L1 Hindi speech dataset with mispronounced speech instances, we utilized the suno/bark-small text-to-speech model \cite{bark-ai}, adhering to the methodology established in prior research on speech synthesis. This process resulted in a dataset comprising over three hours of synthetic Hindi speech generated using ten distinct speaker voices. The dataset contained 1,000 correctly pronounced and 1,000 mispronounced sentence pairs, where three types of phoneme errors---phoneme addition, phoneme deletion, and phoneme modification---were systematically introduced with a probability of 0.05 per phoneme. All audio samples were resampled to 8 kHz with 16-bit encoding and truncated to a duration of 6-8 seconds. Each sample sentence was extracted from authoritative Hindi textbooks \cite{NCERT2015, NCERT2006} and was paired with its mispronounced counterpart, generated using the same speaker embedding to maintain consistency in voice characteristics. Both the correct and incorrect utterances were provided with corresponding transcriptions, and error probability vectors were included for the mispronounced samples, indicating the likelihood of phoneme errors (binary for the training set).\cite{NCERT2006, NCERT2015, NCERT2018}

\begin{table}[t]
    \centering
    \begin{tabular}{|p{3cm}|p{4cm}|}
        \hline
        \rowcolor{gray!20} \textbf{Layer} & \textbf{Hyperparameters} \\ \hline
        \textbf{Conv1D (x5)} & 
        \textbf{Out Channels:} 16 \newline
        \textbf{Dropout:} 0.25\\ \hline
        \textbf{Multi-head Attention} & 
        \textbf{Attention Heads:} 8 \newline
        \textbf{Embedding Dimension:} 64 \\ \hline
        \textbf{Gated Recurrent Unit} & 
        \textbf{Hidden Size:} 64 \newline
        \textbf{Dropout:} 0.2 \\ \hline
        \textbf{Training} & 
        \textbf{Batch Size:} 32 \newline
        \textbf{Sequence Length:} 256 tokens \\\hline
        \textbf{Optimization} & 
        \textbf{Optimizer:} AdamW \newline
        \textbf{Learning Rate:} 3e-6 \\\hline
        \textbf{Regularization} & 
        \textbf{Decoder dropout:} 0.2 \newline
        \textbf{CNN dropout:} 0.25 \\ \hline
    \end{tabular}
    \caption{Hyperparameters for Model Training}
    \label{tab:hyperparameters}
\end{table}

\subsection{Model Architecture}
\begin{figure}[t]
    \centering
    \begin{subfigure}[b]{0.2\textwidth}
        \centering
        \includegraphics[width=\textwidth]{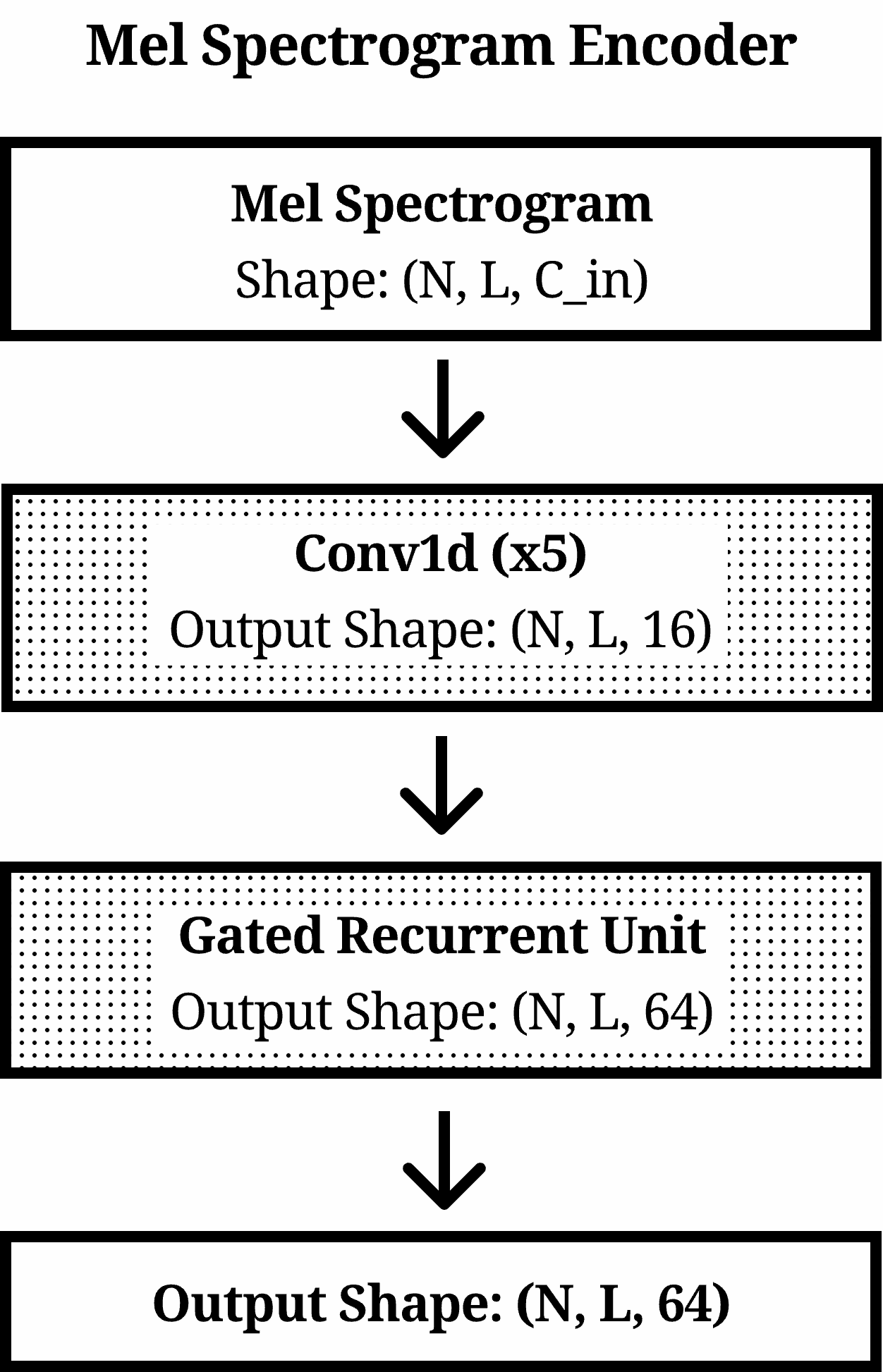}
        \caption*{(a)}
    \end{subfigure}%
    \hspace{0.015\textwidth}%
    \begin{subfigure}[b]{0.2\textwidth}
        \centering
        \includegraphics[width=\textwidth]{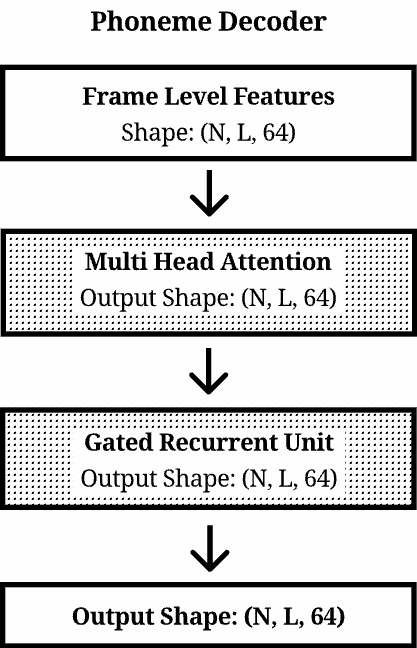}
        \caption*{(b)}
    \end{subfigure}
    \caption{(a) R-CNN Encoder processes input mel-spectrograms with Conv1D layers and a Gated Recurrent Unit (GRU), producing frame-level features, while(b) A-RNN Phoneme Decoder applies Multi-Head Attention on these features and uses another GRU to generate phoneme-level outputs.}
    \vspace{0pt}
    \label{fig:model_architecture}
\end{figure}

We introduce a novel Phoneme Recognition Network tailored for Hindi, building upon the WEAKLY-S model (Fig. \ref{fig:dhvani-model-and-feedback}a). Our architecture features a Recurrent-Convolutional Neural Network (R-CNN) mel-spectrogram encoder for extracting detailed frame-level representations from audio features (Fig.~\ref{fig:model_architecture}a). These representations are then processed by an Attention-based Recurrent Neural Network (A-RNN) phoneme decoder to produce phoneme-level outputs (Fig.~\ref{fig:model_architecture}b). The resulting phoneme-level features are fed to a classification head with 67 units, corresponding to 67 distinct output tokens. Among these tokens, 64 are allocated for Hindi phonemes, with the remaining tokens reserved for end-of-word, end-of-sentence, and padding markers. A comprehensive list of the 64 Hindi phonemes identified by our network along with the source code of the architecture is available on our GitHub repository\footnote{\url{https://github.com/dhvaniapp/Dhvani-187k}}.

\subsection{Feedback Mechanism}
This study presents a comprehensive phonetic personalized feedback system for Hindi pronunciation. It covers the full range of Hindi phonology, including vowels (short and long), consonants (plosives, fricatives, affricates, nasals, approximants, and flaps), diphthongs, and diacritical marks. For each phoneme, the system provides detailed and personalized articulatory guidance, explaining tongue placement, lip shape, airflow, and vocal cord involvement. Specific attention is given to challenging aspects for non-native speakers, such as retroflex and aspirated consonants, as well as nasalization. 

To enhance understanding, visual aids like tongue diagrams accompany the verbal feedback (Fig.~\ref{fig:dhvani-model-and-feedback}b), offering sagittal cross-sections of the oral cavity that depict precise tongue positioning for each phoneme. This multi-modal approach is especially useful in illustrating subtle differences, such as the contrast between dental and retroflex sounds. In addition, aspirated consonants like \textit{dha} and nasalized sounds are broken down to explain their unique articulatory features. Combining structured verbal and visual feedback, the system provides an accessible method for mastering Hindi pronunciation. This approach is valuable for both native and non-native speakers and holds potential for applications in automated language learning and speech recognition technologies.


\section{Evaluation}

\subsection{Model Performance}
To ensure robust evaluation, we employed a rigorous 5-fold cross-validation methodology on our dataset, which comprises 95.05 hours of training data and 5.55 hours of test data. The dataset was carefully partitioned to maintain speaker independence across folds, with 59 distinct speakers in the training set and 19 in the test set, totaling 4,506 and 386 unique sentences respectively. This strategic split ensures that our model's performance metrics reflect its generalization capability across different speakers and phonetic variations. Each fold was evaluated independently, with model parameters initialized randomly to avoid any cross-fold contamination. The reported performance metrics represent the average across all folds, with a standard deviation of less than 2\% for all metrics, demonstrating the stability and reliability of our approach. The specific hyperparameters for each component of our model are detailed in Table~\ref{tab:hyperparameters}.

To assess pronunciation errors, we compare the predicted phonemes with the canonical phonemes and determine the severity of errors based on the confidence of the predicted phoneme error. Leveraging Hindi's phonetic nature, our system can effectively detect pronunciation errors without requiring a dedicated mispronunciation detection network, unlike the original WEAKLY-S model. This approach is made possible by the high correlation between phonemes and words in Hindi.

Table~\ref{tab:literature} shows \textit{Dhvani}'s performance against state-of-the-art CAPT systems, demonstrating substantial improvement and setting a new benchmark for Hindi pronunciation training.


\subsection{Human Evaluation of the Feedback Mechanism}
We conducted an experiment to analyze the effectiveness of our feedback system based on self-evaluation before and after the intervention for 13 different phonemes, using a 5-point Likert scale. These phonemes were chosen based on their difficulty in second language acquisition for Hindi. The experiment consisted of 22 non-native Hindi speakers (12 male, 10 female) from various linguistic backgrounds, aged between 15 and 49 years (mean = 23.29, SD = 8.73). Participants rated their pronunciation of each phoneme before and after using our system.

\begin{figure}
    \centering
    \includegraphics[width=0.5\textwidth]{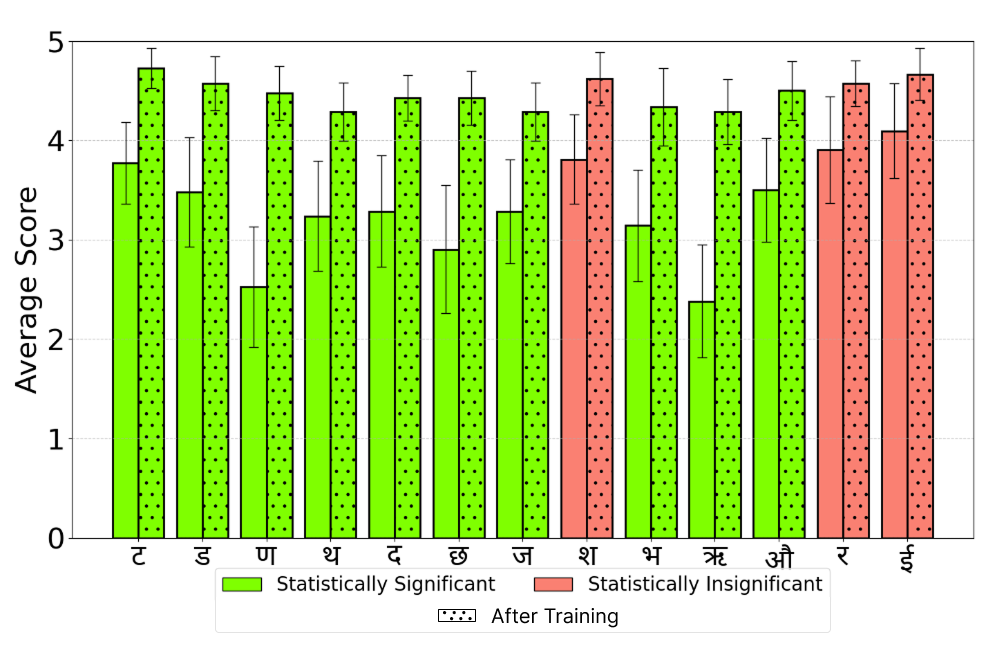}
    \caption{Comparison between pronunciation results before and after using the \textit{Dhvani}'s feedback system for 13 challenging Hindi phonemes using a 5-point Likert scale (n=22 non-native speakers). Box plots show score distributions, with green indicating statistically significant gains for 10 of 13 phonemes after using \textit{Dhvani}'s feedback system. (p $<$ 0.05, Wilcoxon Signed-rank Test) }
    \label{fig:score-delta}
\end{figure}

We used the Wilcoxon Signed-rank Test for testing the statistical significance of pre- and post-intervention response for each phoneme along with Pratt's method of zeros as the data distribution is not normal. A statistically significant p-value of 0.0022 was observed, indicating a meaningful improvement in pronunciation scores. Remarkably, a statistically significant improvement was observed for the pronunciation of 10 out of 13 phonemes after using our system, including dental and palatal consonants (Fig. ~\ref{fig:score-delta}).

Table \ref{tab:survey_results} highlights the user feedback for \textit{Dhvani}. Some users suggested adopting an animated feedback video for better clarity and including a difficulty scale for phonemes. Others also implied that there might be a bias towards North Indian speakers, who comprise a majority of the Hindi-speaking population in India. In general, participants rated the system highly for its effective and personalized feedback and expressed strong confidence in its benefits for new language learners.

\begin{table}[t]
    \centering
    \renewcommand{\arraystretch}{1.2}
    \begin{tabular}{|p{5.55cm}|p{1.6cm}|}
        \hline
        \rowcolor{gray!20} \textbf{Question} & \textbf{Mean $\pm$ SD} \\
        \hline
        Prior awareness of pronunciation techniques & 2.24 $\pm$ 0.81 \\
        \hline
        Helpfulness in understanding pronunciation & 4.00 $\pm$ 0.69 \\
        \hline
        Expected benefit for novice learners & 4.19 $\pm$ 0.85 \\
        \hline
        Helpfulness of explanation detail & 4.24 $\pm$ 0.61 \\
        \hline
        Perceived effectiveness of feedback & 4.24 $\pm$ 0.68 \\
        \hline
    \end{tabular}
    \caption{Survey responses on 5-point Likert scale (n=22)}
    \label{tab:survey_results}
\end{table}

\section{Conclusion and Future Work}
\label{sec:conclusion}
Our research presents a promising CAPT system for Hindi, fulfilling a critical gap in language learning technology for Indian languages. By carefully adapting to Hindi's distinctive phonetic features, our system significantly improves upon existing CAPT technologies. Our approach demonstrates remarkable progress in identifying pronunciation errors and offering constructive and personalized feedback. These improvements are particularly significant given the complexities of Hindi pronunciation. The system is designed not only to enhance accuracy but also to increase learner confidence---a crucial factor in language acquisition. Precise, actionable feedback empowers learners to refine their pronunciation skills effectively. The implications of this research extend beyond Hindi, potentially revolutionizing CAPT for other Indian languages and transforming language education across the Indian subcontinent. Our work aims to broaden access to high-quality speech training, supporting learners as they navigate India's rich linguistic diversity. Overall, this work represents a significant step in addressing the diverse needs of individual learners while contributing to the broader goal of bridging linguistic gaps through innovative speech technology.
\clearpage

\section{Acknowledgments}
The authors would like to thank Harish and Bina Shah School of AI \& Computer Science at Plaksha University for providing seed financial support for this research.

\bibliographystyle{IEEEtran}
\bibliography{refs}

\end{document}